\documentclass[12pt,a4paper]{article}

\usepackage[T1]{fontenc}
\usepackage{microtype}
\usepackage[hmarginratio=1:1,top=32mm,columnsep=20pt]{geometry}

\usepackage{color}
\definecolor{dark-gray}{gray}{0.20}
\definecolor{gray}{gray}{0.30}
\definecolor{light-gray}{gray}{0.80}
\definecolor{dark-red}{rgb}{0.7,0,0}
\definecolor{dark-green}{rgb}{0.1,0.4,0}
\definecolor{dark-blue}{rgb}{0.3,0.3,0.7}
\definecolor{light-blue}{rgb}{0.8,0.8,1}

\usepackage[super,sort&compress]{natbib}
\bibpunct{[}{]}{,\!}{n}{}{}
\usepackage{hyperref}
\usepackage{setspace}
\hypersetup{
	colorlinks=true,
	linkcolor=dark-blue,
	citecolor=dark-red,
	urlcolor=dark-green,
	linktoc=page
}

  \usepackage{a4wide}
  \usepackage{latexsym}
  \usepackage{epsf}
  \usepackage{bbm}
  \usepackage{graphicx}
  \usepackage{amsmath,amssymb,amsthm}
  \usepackage{verbatim}

\newcommand{\f}[2]{\frac{#1}{#2}}

\newcommand{\dd}{\mathrm{d}}

\newcommand{\e}{\mathrm{e}}
\newcommand{\w}{\wedge}
\newcommand{\bbm}{\left(\begin{matrix}}
\newcommand{\ebm}{\end{matrix}\right)}
\newcommand{\bea}{\begin{eqnarray}}
\newcommand{\eea}{\end{eqnarray}}
\newcommand{\be}{\begin{equation}}
\newcommand{\ee}{\end{equation}}
\newcommand{\nn}{\nonumber}

\newcommand{\vol}{\text{vol}}

\renewcommand{\cal}[1]{\mathcal{#1}}

\renewcommand{\d}{\textrm{d}}
\newcommand{\Adt}{\overline{\text{D3}}}
\newcommand{\Ads}{\overline{\text{D6}}}
\newcommand{\Adv}{\overline{\text{D5}}}

\begin{document}
\numberwithin{equation}{section}

\begin{center}

{\LARGE {\bf Unstoppable brane-flux decay of $\overline{\text{D6}}$ branes}}  \\

\vspace{1.5 cm} {\large  U.H. Danielsson$^a$, F. F. Gautason$^{b}$ and T. Van
Riet$^c$ }\\
\vspace{0.5 cm}  \vspace{.15 cm} {${}^a$Institutionen f{\"o}r fysik och astronomi,\\ Uppsala Universitet, Uppsala, Sweden \\[0.3cm] ${}^b$Institut de Physique Th\'eorique, Universit\'e Paris Saclay, CEA, CNRS,\\
\normalsize Orme des Merisiers, F-91191 Gif-sur-Yvette, France\\[0.3cm] 
${}^c$Instituut voor Theoretische Fysica, K.U. Leuven,\\
Celestijnenlaan 200D B-3001 Leuven, Belgium
}

\vspace{0.7cm} {\small \upshape\ttfamily  ulf.danielsson @ physics.uu.se,\\ fridrikfreyr @ gmail.com, thomas.vanriet @ fys.kuleuven.be
 }  \\

\vspace{2cm}

{\bf Abstract}
\end{center}

{\small We investigate $p$ $\overline{\text{D6}}$ branes inside a flux throat that carries $K \times M$ D6 charges with $K$ the 3-form flux quantum and $M$ the Romans mass. We find that within the calculable supergravity regime where $g_s p$ is large, the $\Ads$ branes annihilate immediately against the fluxes despite the existence of a  metastable state at small $p/M$ in the probe approximation. The crucial property that causes this naive conflict with effective field theory is a singularity in the 3-form flux, which we cut off at string scale. Our result explains the absence of regular solutions at finite temperature and suggests there should be a smooth time-dependent solution. We also discuss the qualitative differences between $\overline{\text{D6}}$ branes and $\overline{\text{D3}}$ branes, which makes it a priori not obvious to conclude the same instability for $\overline{\text{D3}}$ branes. }

\setcounter{tocdepth}{2}
\newpage
\tableofcontents
\newpage

\section{Introduction}

A useful way to break supersymmetry in flux backgrounds is to insert branes that preserve different supercharges than the background fluxes and branes. Supersymmetry breaking by combining branes and antibranes is perturbatively unstable since the branes are mobile and can move towards each other and annihilate. 
The virtue of combining branes with fluxes that have opposite orientation is that there is no direct analog of brane/antibrane annihilation. Instead there can be brane-flux annihilation \cite{Kachru:2002gs} which proceeds via first nucleating branes out of the fluxes which then annihilate with the antibranes. The question then is whether this type of supersymmetry breaking can be metastable. If so, then it would be of great practical use in constructing string models for dS vacua \cite{Kachru:2003aw}, inflation \cite{Kachru:2003sx}, near-extremal black hole micro-states \cite{Bena:2011fc, Bena:2012zi} and holographic duals to dynamical supersymmetry-breaking \cite{Maldacena:2001pb, Kachru:2002gs, Argurio:2007qk, DeWolfe:2008zy,  Klebanov:2010qs, Kutasov:2012rv, Bertolini:2015hua} (and see \cite{Retolaza:2015nvh} for related work).

In \cite{Kachru:2002gs} it was indeed argued that metastable states do exist for $p$ $\overline{\text{D3}}$ branes in the Klebanov--Strassler throat \cite{Klebanov:2000hb} with three-form fluxes carrying $M\times K$ D3 charges, where $K$ and $M$ are respectively the $H_3$ and $F_3$ flux quanta.   Brane-flux annihilation proceeds via the Myers effect \cite{Myers:1999ps}: the $p$ $\overline{\text{D3}}$ branes polarise into an NS5 brane wrapping a two-cycle inside the 3-cycle (i.e. A-cycle) at the tip of the throat. The NS5 brane carries $\Adt$-charge that is a function of its position on the A-cycle. When the NS5 brane pinches off at the opposite side of the 3-cycle it induces $M-p$ D3 charges instead of $p$ $\Adt$ charges. Effectively, the motion over the 3-cycle corresponds to brane-flux annihilation: out of the background fluxes $M$ D3 branes have materialised and at the same time $K$ dropped by one unit. The $M$ D3 branes then annihilate with $p$ $\Adt$ branes to give a supersymmetric vacuum with $M-p$ D3-branes. Kachru, Pearson and Verlinde (KPV) argued \cite{Kachru:2002gs} that this supersymmetry breaking is metastable in the limit of small $p/M$. This limit is the same limit in which the characteristic size of 3-sphere at the tip $R^2_{S_3}\sim g_S M\ell_s^2$ is much bigger then the characteristic size of the $\Adt$ horizon $R^4_{\Adt}\sim g_s p\ell_s^2$. Hence one might expect the $\Adt$ to only modify the geometry locally. KPV argued the metastability by computing the potential energy for the NS5 brane and finding it has a local minimum at a radius $r\sim (p/M)\sqrt{g_sM}\ell_s$.  In the past years concerns have been raised in the literature about the validity of this result. The first concern is that the NS5 potential was computed by using an action obtained from S-dualising the D5 action. The D5 action is valid at weak coupling and hence the obtained NS5 action can only be argued to be valid at strong coupling. An alternative analysis using the non-Abelian D3 action also suffers from the same problem since an S-dual version of the standard D3 action is required. A second concern is that the KPV computation works at the probe level and backreaction effects should therefore be subleading and tunably small. However many recent works have shown that the backreaction leads to a singularity which is difficult to interpret. 

In this paper we study whether backreaction destroys the metastable state of $\Ads$ branes. For that we must first establish the metastability of $\Ads$-branes at the probe level which has not appeared in the literature before (see however \cite{Gautason:2015tla}). We then indeed find that backreaction destabilises the $\Ads$ branes. We are unable to address the $\Adt$-brane stability conclusively but we comment on it.

In section  \ref{back} we recall the existing results on the backreaction of antibranes. Section 3 contains a description of the background flux solution in which we place the $\Ads$-branes.  We then show, in section \ref{probe}, that $\Ads$-branes give rise to metastable states at the probe level when $p/M$ is small ($M$ is the Romans mass, $p$ is the antibrane charge). This meta-stable state is however washed away when backreaction corrections are added (assuming $g_sp$ is large enough to allow for a supergravity description) as shown in section \ref{sec:backreact}. In section \ref{sec:p<6} we comment on the case of branes of lower dimensionality, and discuss a possible loophole that could allow the interesting case of $\Adt$ branes to escape this instability. We conclude in section \ref{concl}. All explicit computations can be found in the Appendices.

\section{Backreaction effects and instabilities}\label{back}
An explicit study of antibrane supersymmetry breaking is hard due to the lack of explicit solutions in the cases of interest. The best understood solutions at the moment are based on antibranes smeared over the A-cycle  which has been most studied for $\Adt$ branes in the KS throat \cite{Bena:2009xk} (see also \cite{McGuirk:2009xx}). The only case for which a quantitative analysis is possible that does not rely on smearing are $\Ads$-branes since those solutions are described by ODE's\cite{Blaback:2011pn}\footnote{One exception seems to be $\overline{\text{M2}}$ branes in singular geometries \cite{Giecold:2013pza, Cottrell:2013asa}.}. The qualitative feature that these studies have shown is the presence of a singularity in the $H_3$ flux density. This property can be shown to be unavoidable even for localised branes \cite{Gautason:2013zw, Blaback:2014tfa}. Two prominent interpretations of this singularity have appeared in the literature in the past years.
\begin{itemize}
\item Singularities are actually to be expected \cite{Dymarsky:2011pm} since near the $\Adt$-brane we have an $AdS_5\times S^5$ perturbed by three-form fluxes. Analogous, albeit supersymmetric, examples of this type are known to give rise to singularities that are resolved by letting the 3-branes \emph{polarise} to ($p,q$)-5 branes \cite{Polchinski:2000uf}. For smeared $\Adt$-branes there is a growing body of work that indicates that certain expected polarisation channels seem absent \cite{Bena:2012tx, Bena:2012vz, Gautason:2015ola} and others lead to tachyonic modes \cite{Bena:2014jaa, Bena:2014bxa, Danielsson:2015eqa, Bena:2016fqp}. Analogous results for fully localized branes are not as strong and a recent paper found that polarised $\Adt$-branes can be made regular by fixing boundary values of some fields \cite{Cohen-Maldonado:2015ssa}. It still remains to show that a full solution with these boundary values exists. 
\item The 3-form singularity of $\overline{\text{D}p}$ solutions has a simple interpretation \cite{Blaback:2011nz} which can be found in the sign of the divergent (but integrable) charge density $F_{6-p}\wedge  H_3$. This sign is opposite to the sign of the antibrane charge. The reason is that the flux is attracted gravitationally \emph{and} electromagnetically towards the antibranes. If the antibranes are replaced instead with branes, the electromagnetic repulsion will exactly counterbalance the gravitational pull \cite{Blaback:2010sj}. This entails an obvious instability since a clumping of fluxes carrying D$p$ charges will enhance the brane-flux annihilation. This can be seen at the level of the probe actions describing the Myers effect \cite{Blaback:2012nf, Danielsson:2014yga, Gautason:2015ola} or from the point of view of bubble nucleation \cite{Danielsson:2014yga}. If the flux clumps in an unbound way one necessarily crosses a critical value that causes immediate brane-flux decay. If this picture is correct there should exist a smooth time-dependent solution describing the clumping of flux, which reaches a critical value at which the antibranes decay and then the flux clumping should stop and the BPS state should be reached after the closed string radiation decouples.  This picture is not inconsistent with the expectation that backreaction is \emph{small} when $p/M$ is small, but the essential property that makes this possible is that the \emph{local} backreaction, which can be big, determines stability. 
\end{itemize}
In this paper we demonstrate that the second viewpoint is correct for $\Ads$ branes at large $g_sp$ but still arbitrarily small $p/M$.

\section{$\overline{\text{D6}}$ brane solutions}\label{solution}

In order to analyse  brane-flux annihilation, we first require
a background with fluxes that carry D6 charges without the presence of localized D6 branes. This is possible in massive IIA supergravity as can be seen from the Bianchi identity for $F_2$:
\be
\d F_2 = F_0 H_3 + N\delta_6\,.
\ee  
Here $F_2$ is the RR 2-form field strength, $F_0$ is Romans mass and $H_3$ is the NSNS 3-form field strength. Notice that the term $F_0 H_3$ appears on the same footing as the D6 source term $N\delta_6$. Clearly when $F_0 \neq 0$ this term acts as a smooth source for 6-brane charge.

A general study of backgrounds with these ingredients was carried out in \cite{Blaback:2011pn}. It includes  the ``massive D6'' brane with flat worldvolume and non-compact transverse space \cite{Janssen:1999sa} and D6 branes with AdS$_7$ worldvolume and compact transverse space \cite{Blaback:2011pn, Apruzzi:2013yva}. We start with brane-flux decay in the first example but before doing so we emphasize why the $\Ads$-brane is special.

\subsection{Why the $\overline{\text{D6}}$ is special}
$\Ads$ branes stand out against antibranes of other dimensionality for their simplicity. This is most obvious in the description of the backreaction of $\Ads$ branes, which is described by ODE's \cite{Blaback:2011nz,Blaback:2011pn,Apruzzi:2013yva} without having to smear the branes. A second feature--crucial for this paper--is that the annihilation of the $H_3$-flux does not proceed via a polarisation into a higher-dimensional object. All $\overline{\text{D}p}$ branes with $p<6$ polarise into an NS5 branes \cite{Gautason:2015tla}. $\Ads$ branes instead polarise into KK5 branes, which are smaller in dimension. However the KK5 assumes a circular isometry \emph{transverse} to its worldvolume, one can therefore think of KK5s as 5-branes that are smeared over a circle. This circular isometry direction will live inside the $\Ads$ worldvolume such that the KK5 brane looks like a 6-brane and its backreaction is identical to that of the $\Ads$ brane since it carries $\Ads$ charge and tension. This is related to the fact that $\Ads$ solution is the only antibrane solution for which the singularity in $H_3$ is consistent with the singularity of the metric. In other words: the backreaction of the diverging $H_3$ flux does not destroy the local $\overline{\text{D6}}$-metric, whereas it would for $p<6$. For instance for $p=3$ one finds that the $H_3^2$ scalar blows up near the horizon, but the horizon of a 3-brane should be smooth $AdS_5\times S^5$. Hence the 3-form fluxes necessarily destroy the local $AdS_5$ throat, which is well known from the Polchinski-Strassler (PS) model \cite{Polchinski:2000uf}. In the PS model the singularity at the would-be horizon can be turned into a physical singularity corresponding to $(p,q)$ 5-branes. Something similar can be expected for the $\overline{\text{D3}}$ solution and indeed reference \cite{Cohen-Maldonado:2015ssa} has shown that a local 5-brane background with good singularities it is at least not inconsistent with having ISD fluxes at the UV and a non-zero (generalised) ADM mass.

\subsection{$\overline{\text{D6}}$ branes in flat space}
We start with a flux background geometry with a flat D6 worldvolume. The metric and dilaton\footnote{We use string frame throughout.} is that of a D6 brane
\bea
\dd s^2 &=& S^{-1/2}(\dd s^2_7) + S^{1/2}[\dd r^2 + r^2 \dd\Omega_2^2]~,\nonumber\\
F_2 &=& -g_s^{-1}\tilde{\star}_3 \dd S~,\nonumber\\
e^{\phi} &=& g_s S^{-3/4} ~.\label{BPSbackground}
\eea
To have D6 charge dissolved in fluxes we include a non-trivial $H_3$ flux and Romans mass given by
\bea
H_3 &=&   (Mg_s/\ell_s) \vol_3~,\nonumber\\
F_0 &=&  M/\ell_s\,.
\eea
Here $\tilde{\star}_3$ is the unwarped Hodge operator on the transverse space such that $\tilde{\star}_3 1 
= r^2 \dd r \w \Omega_2=\vol_3$
and $\Omega_2$ is the volume form on the unit 2-sphere. We denote the line element of 7D Minkowski space by
$\dd s^2_7$. Further on, once we discuss the KK5 brane it will turn out to be useful to pick out a specific direction $\psi$ that plays the role of the isometry direction of the KK5. We  then write $\dd s_7^2 =\dd s^2_6 + \dd \psi^2$ with $\psi$ a circle direction.
The fluxes obey a Hodge-duality relation: 
\be\label{ISD}
H_3 = \e^{\phi} \star_3 F_0~,
\ee
where the Hodge-star $\star$ includes warp factors. This condition is identical to the famous \emph{Imaginary Self-Dual (ISD)} condition for fractional D3 brane backgrounds after performing three T-dualities along the Minkowski directions. One can furthermore verify that (\ref{ISD}) ensures a no-force condition for probe D6 branes. $\overline{\text{D6}}$ branes on the other hand do feel a force.

The Bianchi identity for $F_2$ (with $N$ D6-brane sources
located at the origin $r=0$) implies
\[
\tilde{\triangle}_3 S +  \left(\f{M g_s}{\ell_s}\right)^2  = 0~,
\]
to which a spherically symmetric solution is
\be
S =  v^2 + \frac{g_s \ell_s N}{4\pi r} -\frac{(Mg_s r)^2}{6\ell_s^2}~,
\ee
where $N$ is the number of D6-branes sitting at $r=0$.
We will put $N$ to zero for now and consider a background without any explicit D6 sources. This solution is then regular around $r=0$.

The solution has a peculiar singularity for
\be
r\to \f{\sqrt{6}v\ell_s}{M g_s}~.
\ee
In order to examine the local behaviour of the singularity we expand around it:
\be
r = \f{\sqrt{6}v\ell_s}{M g_s} - \delta r\quad\text{with}\quad (\delta r)^5 =\f{\sqrt{6} \ell_s}{2 v M g_s}\left(\f{5}{4}x\right)^4~, 
\ee
and find
\be
\dd s^2\approx\left(\f{2\sqrt{6} \ell_s}{5v g_s M x}\right)^{2/5}\dd s_7^2 + \dd x^2 + \left(\f{\sqrt{6}v \ell_s}{g_s M}\right)^2\left(\f{5v g_s M x}{2\sqrt{6}\ell_s}\right)^{2/5}\dd\Omega_2^2~.
\ee
This is precisely the local singularity structure of an O6-plane (see appendix \ref{flatO6}). A discussion of this singularity can be found in \cite{Hartnett:2015oda} where it was argued that strings are well behaved in the singular background. Our interpretation of the singularity as an O6-plane confirms that analysis. We will not go into more details on the global analysis of the solution since the computation we perform is local to $r=0$. 

This solution is not supersymmetric. Supersymmetric solutions for D6 branes in massive IIA do not preserve the Poincar\'e symmetries of the D6 worldvolume as shown in \cite{Janssen:1999sa, Imamura:2001cr}. The flux background is however \emph{extremal} in the following sense
\begin{itemize}
\item The expression of the metric and form fields is the one typical for extremal $p$-branes.
\item There is a no-force condition for inserting D6 branes into the background whereas $\overline{\text{D6}}$ branes are pulled towards the tip of the throat.
\item It is T-dual to supersymmetry breaking using (3,0)-fluxes in fractional D3 backgrounds which is rather well understood and does not influence any of the details of brane-flux decay in that context \cite{Kachru:2002gs}. 
\end{itemize} 

\subsubsection*{Inserting $\overline{\text{D6}}$ branes}
To describe the same throat geometry but with a stack of $\overline{\text{D6}}$'s at the tip, we use the following general Ansatz preserving the worldvolume symmetries (possibly broken at finite temperature) and the transverse rotational symmetries \cite{Blaback:2011nz} (in string frame): 
\bea
\dd s^2 &=& \e^{2A}(-e^{-2f}\dd t^2 + \dd s_6^2) + \e^{2B}[e^{-2f}\dd r^2 + r^2 \dd\Omega_2^2]~,\nonumber\\
F_0 &=&  M/\ell_s ~,\nonumber\\
H_3 &=& \lambda\e^{\phi} \star_3 F_0~,\nonumber\\
F_2 &=& \e^{-7A}\star_3\dd \alpha ~.\label{background}
\eea
From combining the form field equations of motion one can deduce that
\be 
\alpha=\lambda\e^{7A-\phi +f}\,.
\ee
Solutions of the above form will break the ISD condition \eqref{ISD} whenever $\lambda \neq 1$ and such backgrounds exert non-zero forces on probe D6 branes. 
The extremal flux background discussed earlier is recovered for the choices $\lambda=1$, $\e^{-4A} = \e^{4B}= S$ and $\e^{\phi} = g_s S^{-3/4}$.

It was shown in \cite{Blaback:2011nz, Blaback:2011pn, Bena:2013hr} that insisting that the metric has an $\overline{\text{D6}}$ singularity at $r=0$ and asymptotes to the ISD solution, we find that $\alpha$ goes to a \emph{non-zero} constant near the $\overline{\text{D6}}$-horizon\footnote{which has zero size at zero temperature.} which we denote $\alpha_0$. This implies:
\be \label{lambdablowup}
\lambda = \alpha_0 e^{-7A+\phi -f} \rightarrow \infty ,
\ee
since the combination $e^{-7A+\phi -f}$ blows up at zero or non-zero temperature. At non-zero temperature this is solely due to $e^f$ going to zero near the horizon whereas the other fields remain finite. Whereas at zero temperature $e^f=1$ but $e^{-7A+\phi}$ is combination that necessarily diverges near a 6-brane.

This singularity is not a coordinate artefact since it appears in scalar quantities like the $H_3$-flux density:
\be
e^{-2\phi}|H_3|^2\rightarrow \infty\,.
\ee
One readily verifies that the singularity  in the associated charge density $M H$ is still integrable, since $H_3\sim r^{-1}\text{Vol}_3 =  r\d r\d\Omega_2$ near the source at $r=0$. Another peculiar feature is the fact that the backreaction of the fluxes does not destroy the local 6-brane geometry (at zero temperature). This is different for the $\overline{\text{Dp}}$ solutions with $p<6$ as explained in the Introduction.

\begin{figure}[h!]
\centering
\includegraphics[width=.65\textwidth]{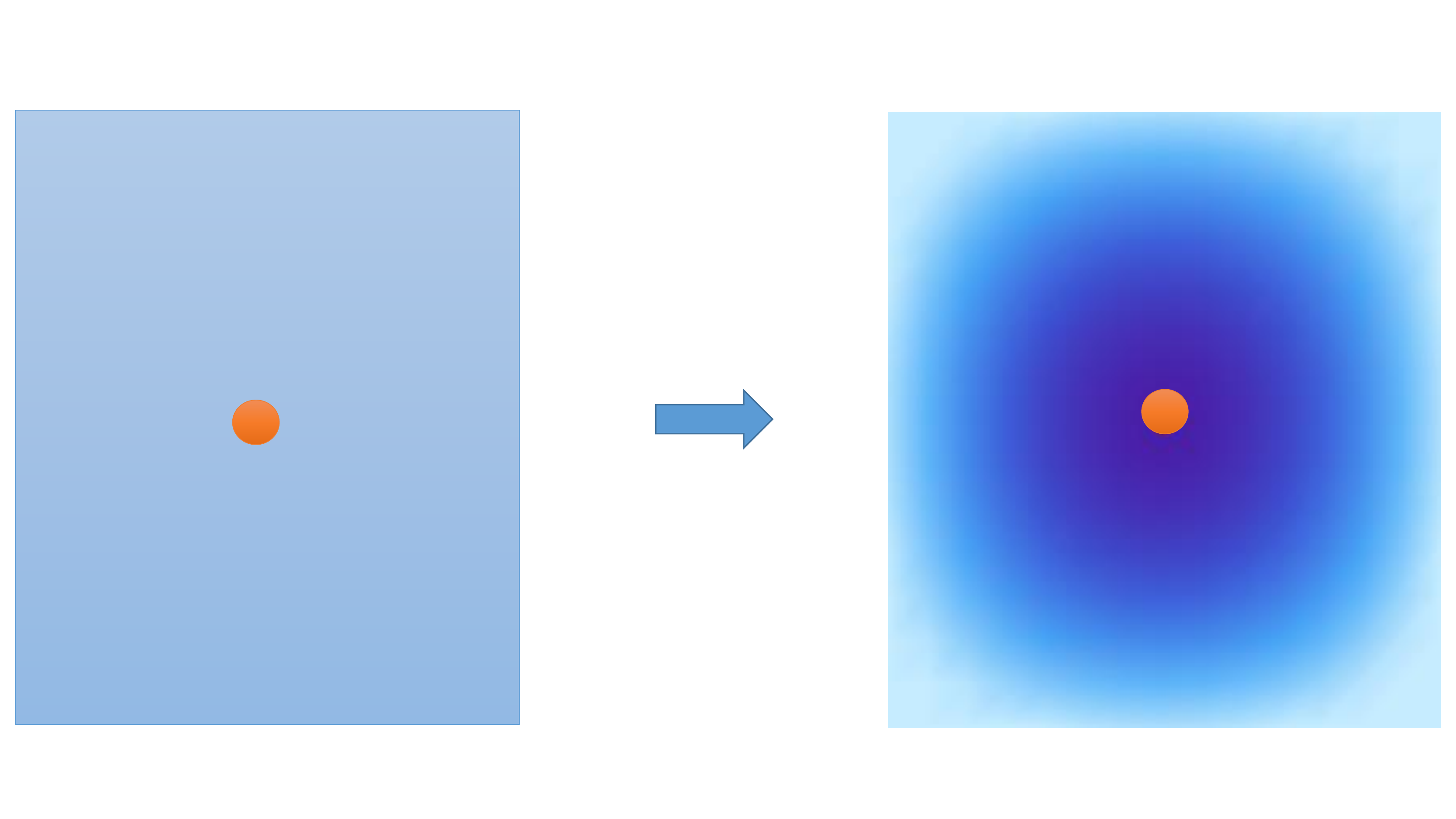}\caption{\small{\emph{
The clumping of positively charged fluxes near a negatively charged antibrane. The blue region corresponds to the flux density and the darker the blue the higher the density.  The red dot represents the antibrane. The left picture illustrates the probe approximation and the right includes backreaction effects.}}}%
\label{Figure1}
\end{figure}

The sign of the singularity in $\lambda$ is such that it corresponds to a diverging $D6$ charge density dissolved in the fluxes  as illustrated in figure \ref{Figure1}.

Given the interpretation of this singularity as a diverging charge density one should be worried about the stability of this background against brane-flux annihilation. Heuristically one expects  a large charge density dissolved in fluxes to increase the probability of localised and mobile D6 branes materialising out of the flux cloud that subsequently annihilate against the $\overline{\text{D6}}$ branes. This is the topic of section \ref{probe}.

\subsection{Warped AdS$_7$ with $\overline{\text{D6}}$ branes}
If one allows the $\Ads$ worldvolume to be AdS$_7$ instead of Mink$_7$ the space transversal to the brane can be made compact and conformal to $S^3$ \cite{Blaback:2010sj}. The resulting solution was described qualitatively in \cite{Blaback:2011nz, Blaback:2011pn}. A much deeper understanding of the solution as well as a better numerical control was reached in \cite{Apruzzi:2013yva}, where it was understood that the solution can even be supersymmetric, rectifying some statements in \cite{Blaback:2011nz}\footnote{ Although non-supersymmetric solutions are possible as well in this setup \cite{Junghans:2014wda, Apruzzi:2016rny}.}. Consequently the dual SCFT's were uncovered in \cite{Gaiotto:2014lca} (see also \cite{Cremonesi:2015bld}).

\begin{figure}[h!]
\centering
\includegraphics[width=.55\textwidth]{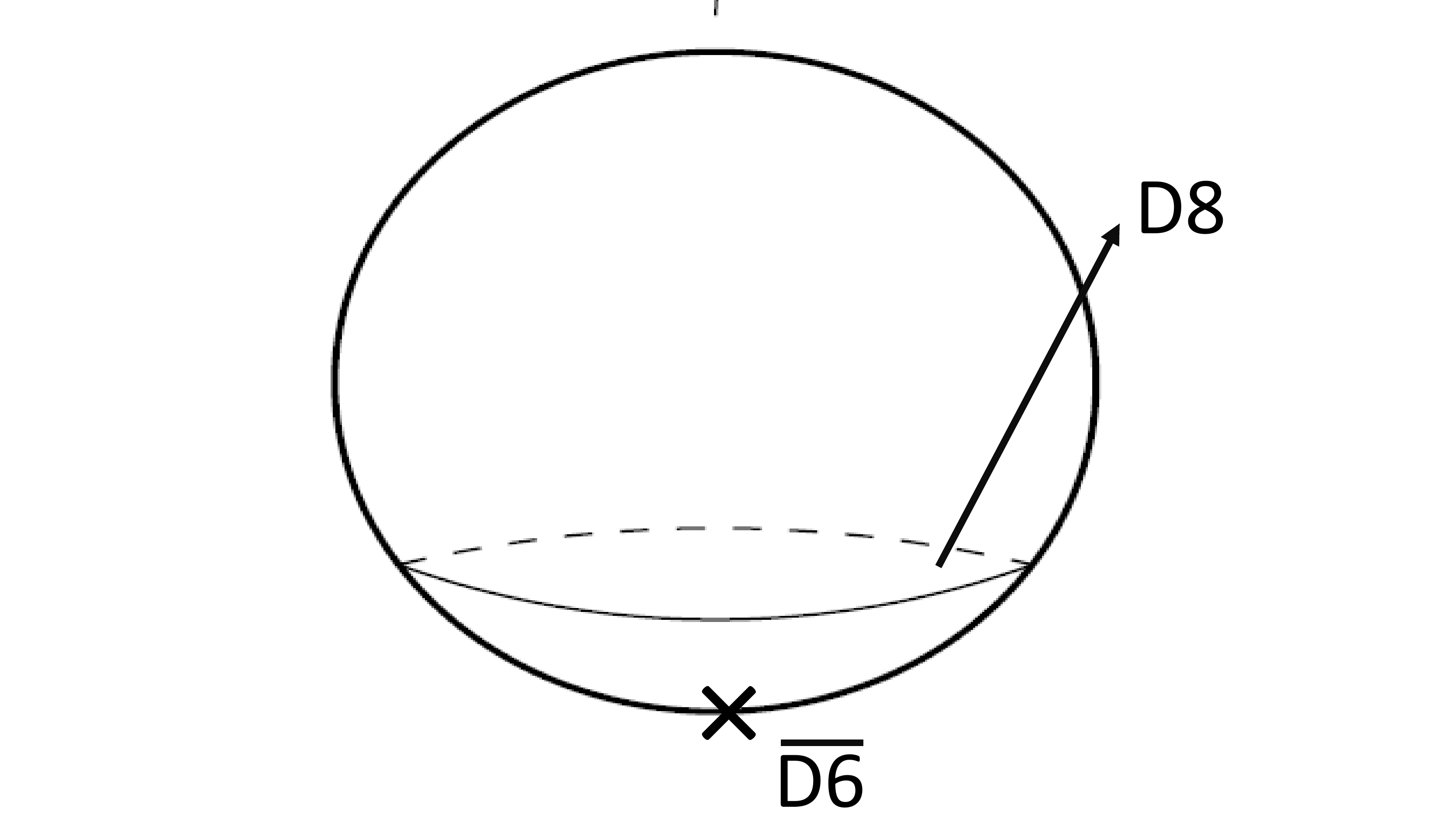}\caption{\small{\emph{$\Ads$ charges can be distributed over the $S^3$ either by point-like $\Ads$ branes or by spherical D8 branes with fluxes wrapping contractible $S^2$'s.}}}%
\label{Figure4}%
\end{figure}

What counts for this paper, all details aside, is that there are various AdS$_7$ solutions. They can be classified according to the way the six-brane charges are divided over the compact manifold. The $\Ads$ charges are either due to explicit $\Ads$ branes sitting at the poles of the $S^3$ or due to spherical D8 branes wrapping contractible $S^2$'s inside the $S^3$ as in figure \ref{Figure4}. Those D8 branes can induce $\Ads$ charges when they have the right kind of worldvolume fluxes on them.

The presence of D8 branes carrying the $\Ads$ charges is a crucial difference with the Minkowski solutions treated above. In the latter case it can be shown that the D8 branes would never sit at a stable position and pinch off to become pure $\Ads$ branes \cite{Bena:2012tx}. Only in AdS$_7$ can they reach a stable position \cite{Junghans:2014wda}. Crucially, when the charges are carried by spherical D8 branes they do not cause flux-clumping singularities \cite{Bena:2012tx}.

The role of the D8 brane can be confusing so we spend a few more words on this. Fluxes can decay against the $p$ $\overline{\text{D6}}$-charges in two ways \cite{Gautason:2015tla}: 1) either by having a D8 brane polarising and move all the way over the ``B''-cycle which decreases the Romans mass or 2) by having a KK5 brane moving inside the D6 worldvolume which decreases the $H_3$-flux.  In this paper we focus on the second channel the D8 will be stuck at a fully stable position in the case of the AdS$_7$ worldvolume or cannot polarise when the worldvolume is Minkowski \cite{Bena:2012tx}. The effect on the KK5 channel should then be such that the KK5 polarisation does not occur at all for AdS worldvolume\footnote{This is stricktly speaking an assumption for our setup but has been verified explicitly in the T-dual picture \cite{Gautason:2015ola}} and, as we argue in this paper, the KK5 will polarise and not even come to stop for flat worldvolumes. This drastic qualitative difference is caused by having the D8 brane polarised in one case and not in the other \cite{Gautason:2015ola}. Because if the D8 polarises it removes the 3-form singularity, which lies at the heart of the immediate brane-flux decay.

\section{Brane-flux annihilation at probe level}\label{probe}

\subsection{Brane-flux decay and T-duality}

To analyse the stability of $\Ads$-branes in the background in question we  first study the probe approximation where the branes do not backreact on the geometry. We must identify branes that can carry the $\Ads$-charge away, similar to the spherical NS5-brane in KPV \cite{Kachru:2002gs}. These turn out to be KK5-branes that source D6-charge on their worldvolume \cite{Gautason:2015tla}. This may seem strange as the KK5 is a 5-brane whereas the $\Ads$ is a 6-brane, but the KK5 has one special transverse direction, the NUT direction, which assumes circular isometry. For all intents and purposes we can think of the KK5 as a brane that is smeared along this NUT direction and hence can carry $\Ads$ charge. In our setup the NUT direction of the KK5-branes is a circle in the $\Ads$ worldvolume parametrized by $\psi$.
In order to not introduce monopole KK5 charges, we consider a pair of KK5 and $\overline{\text{KK5}}$ that share a NUT direction $\psi$ but are otherwise separated on that circle. We illustrate the $\Ads$ decay process in figure \ref{decay}:
\begin{figure}[h!]\label{decay}
\centering
\includegraphics[width=.75\textwidth]{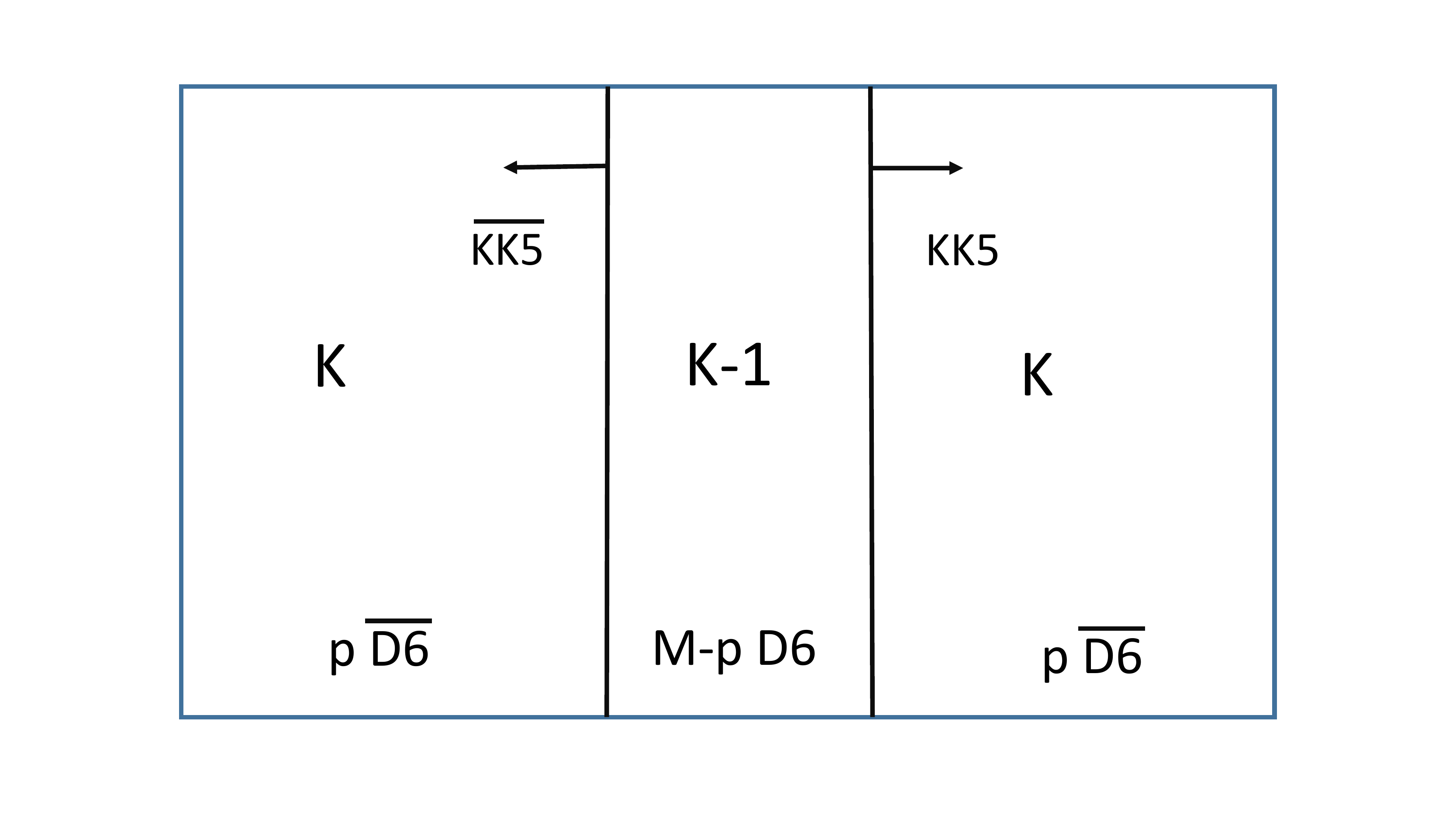}\caption{\small{\emph{The decay of $p$ $\Ads$-branes to $M-p$ D6-branes. The picture represents the worldvolume directions of the six-branes where the horizontal line is a circle. Inside the 7D worldvolume a KK5/$\overline{\text{KK5}}$ pair nucleates and moves outward on the circle to meet at the other side. Inside the pair the $H_3$-flux $K$ drops one unit and the $p$ $\Ads$-branes are replaced by $M-p$ D6-branes.}}}
\end{figure}
The  spacetime domain wall that mediates the decay of $H$-flux and $\Ads$-charge is a NS5-brane wrapping the $S^1$ but traveling radially in Mink$_6$.

\subsection{The probe potential}
To compute the probe potential for KK5 branes in massive IIA we could use the actions derived in \cite{Eyras:1998hn}. It is however technically easier to work in the T-dual frame, and use that the probe  potentials are invariant under T-duality \cite{Gautason:2015tla}. The T-dual setup involves a pair of probe NS5-branes, with opposite charge in the geometry Mink$_6\times S^1 \times M_3$ where the branes are localised on the circle and on $M_3$. The pair of NS5-branes induce $\Adv$ charge between them as explained below. This calculation was already sketched out in \cite{Gautason:2015tla} but we will go through the details here for completeness.

We start by writing down the background T-dual to \eqref{BPSbackground} along the $\psi$-direction,
\bea
\dd s^2&=& S^{-1/2}\dd s_6^2+ S^{1/2} \left(\dd \psi^2 + \dd r^2 + r^2 \dd \Omega_2^2\right)~,\nn\\
F_1 &=& (M/\ell_s) \dd \psi~,\nn\\
F_3 &=& g_s^{-1}\tilde\star_4 \dd S~,\nn\\
H_3 &=&  (Mg_s/\ell_s) \tilde\star_4 \dd \psi~,\nn\\
\e^{\phi}&=& g_s S^{-1/2}~.
\eea
The function $S$ takes takes the same form as above
\be\label{Seq}
S = v^2 - \f{(g_s M r)^2}{6 \ell_s^2}~.
\ee
After T-duality the coordinate $\psi$ remains periodic with $\psi\sim \psi+\ell_s$.
Before continuing on to the probe computation we review how a NS5-$\overline{\text{NS5}}$ pair can induce a $\overline{\text{D5}}$-brane charge. Let the NS5-brane be located at $\psi=\psi_0$ and the $\overline{\text{NS5}}$-brane be located at $\psi = -\psi_0$. The probe action relevant for these branes is
\be\label{braneaction}
S = -\mu_5 \int \e^{-2\phi}\sqrt{-g}\sqrt{1+\e^{2\phi}{\cal G}_\pm^2} \pm \int(B_6 - {\cal G}_\pm C_6)~.
\ee
The upper sign is used for the NS5-brane and the lower is for the antibrane. The field ${\cal G}_\pm$ is a worldvolume field
\be
{\cal G}_\pm = \pm \f{p}{2} - C_0~,
\ee
where $p$ is a constant and $C_0$ is the local gauge potential for $F_1$. The D5 charge induced by the brane pair can be inferred from the sum of the above WZ couplings,
\be
-\mu_5 \int \left[\left(\f{p}{2}-C_0\right)C_6|_{\psi=\psi_0} - \left(-\f{p}{2}-C_0\right)C_6|_{\psi=-\psi_0}\right] = -\mu_5\left(p-\f{2M\psi_0}{\ell_s}\right)\int C_6~,
\ee
where we have used $C_0 = (M/\ell_s)\psi$ and that $C_6$ is independent of $\psi$. The induced D5-charge is therefore controlled by the combination $p-2M\psi_0$. Since $\Delta\psi\equiv 2\psi_0$ is the separation of the brane pair we see that when pulled together to the point $\psi =0$ they induce the charge of $p$ $\overline{\text{D5}}$ whereas on the other pole $\psi_0=\ell_s/2$ the charge is $p-M$. The D5-brane charge carried by the NS5-$\overline{\text{NS5}}$ pair depends on the relative seperation of the branes.

\begin{figure}\label{fig:potential}
\centering
\includegraphics[width=0.7\textwidth]{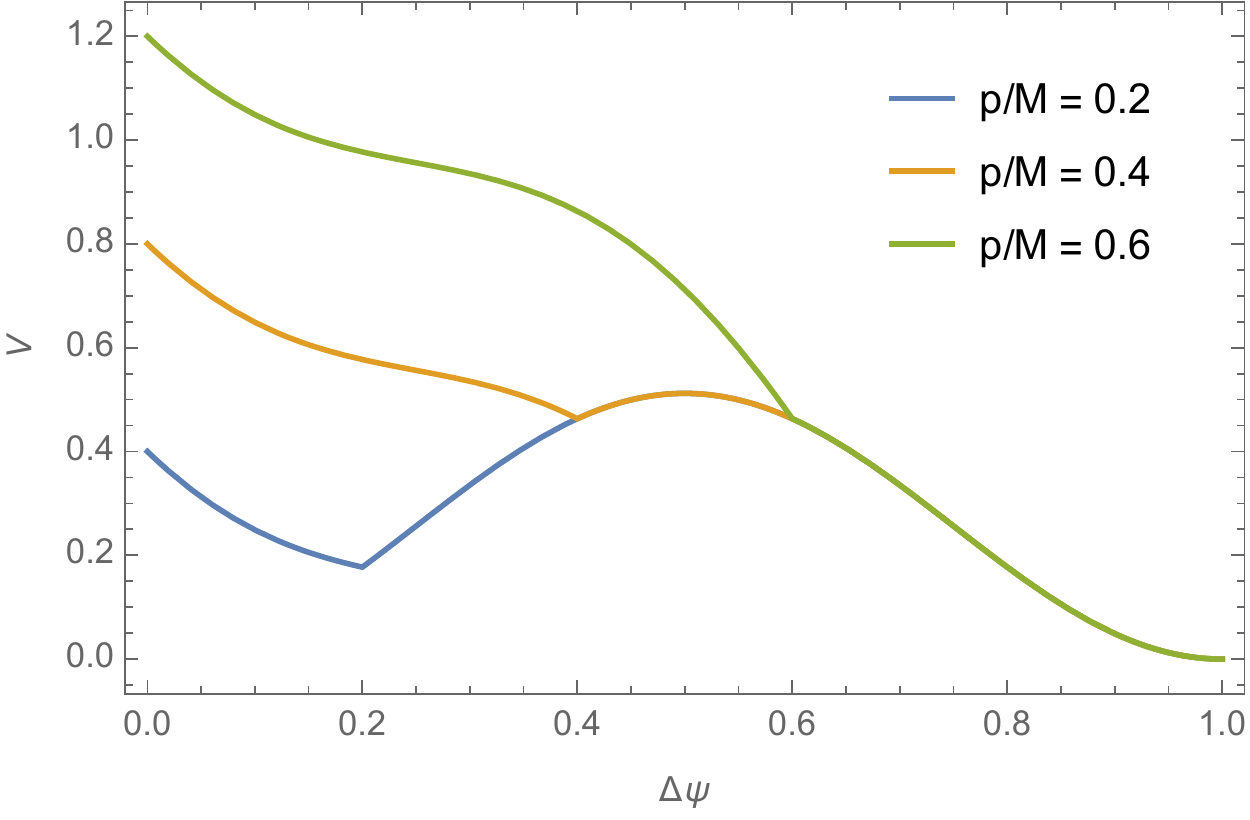}
\caption{\small{\emph{Probe potential with $A=1$ and $M = 10$, $g_s = 10^{-3}$. The values for $p/M$ are indicated in the figure and are displayed with different colors.}}}
\end{figure}

The computation of the probe potential is spelled out in Appendices \ref{App1:potential} and \ref{App2:potential}. Including the interaction term in the full probe potential we find:
\bea
V &=& \mu_5 g_s^{-1}M v^{-2}\left(\left|\f{p}{M} - \f{\Delta\psi}{\ell_s} \right|-\f{\Delta\psi}{\ell_s} +\f{p}{M}
+ \f{2}{ Mg_s}  \f{\sin^2 \left(\pi \Delta \psi/\ell_s\right)}{4\pi^4 + \sin^2 \left(\pi \Delta \psi/\ell_s\right)}\right) ~.\label{probeV1}
\eea
The potential is displayed in figure \ref{fig:potential} and shows that the hight of the barrier is controlled by $M g_s$. 
One can verify that there is a metastable state for $M g_s$ large whenever
\be
\f{p}{M} < 0.5~.
\ee

\section{Including backreaction}\label{sec:backreact}

\subsection{Effective field theory approaches}
A probe result should be thought off as a leading-order contribution in a perturbative series. If the series would be an expansion in $p/M$ one can trust the leading order result in the limit of vanishing $p/M$. For the special case $p=1$ some initial steps towards a direct effective field theory approach were taken in \cite{Michel:2014lva} based on \cite{Goldberger:2001tn}. An alternative description (see for instance \cite{Polchinski:2000uf}) is the one we take in this paper where $p/M$ is small but $g_s p$ is nevertheless big. 

An effective field theory description of any kind will have to deal with the singularities typical to brane solutions, such as the singular warpfactors and diverging gauge fields sourced by the brane. Just as in standard electrodynamics one does not expect diverging field strengths to cause any harm as they correspond to an infinite self-energy that is effectively subtracted off in the probe action. For diverging warpfactor, the same story applies. 

\begin{figure}[h!]
\centering
\includegraphics[width=.65\textwidth]{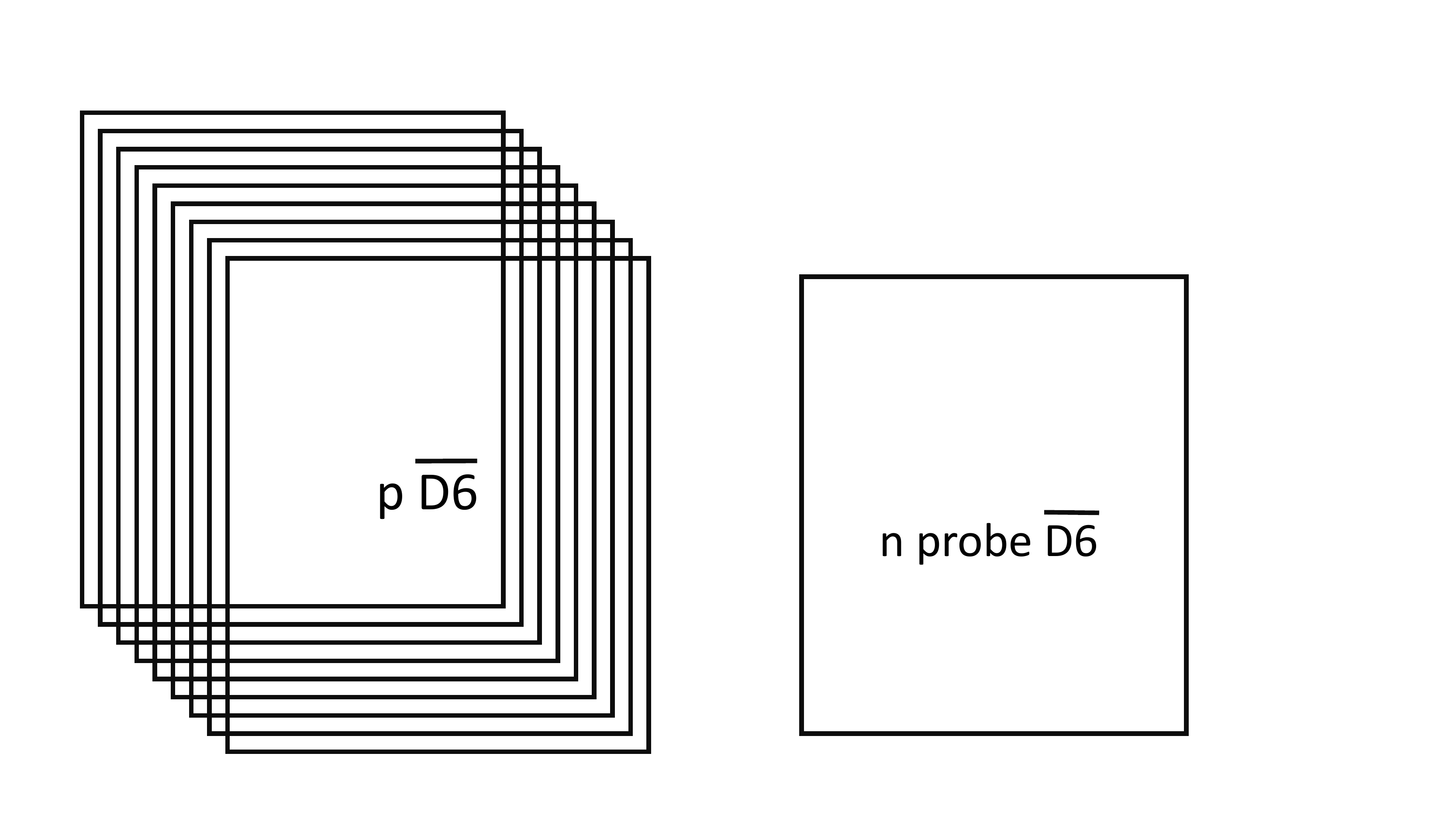}\caption{\emph{Out of a stack of $p$ backreacting branes $n$ branes are pealed off and considered as probes in that background when $n<<p$.}}%
\label{stack}%
\end{figure}

The diverging $H_3$ field on the other hand is not as clear since it is not sourced by the 6-brane directly. It is tempting to think of it as the diverging field strength corresponding to the NS5(KK5)/$\Ads$ boundstate since NS5-branes do source singular $H_3$ flux\footnote{Similarly for KK5-branes in massive IIA.}. However it has been found in \cite{Cohen-Maldonado:2015ssa, Cohen-Maldonado:2016cjh} that this cannot be, at least not within the Ansatz presented in this paper, since then the divergence in $H_3$ is not of the right kind to correspond to a NS5 (KK5) source. As mentioned earlier the singularity really corresponds to a divergence in the flux clumping, which makes the D6 charge density dissolved in flux diverge near the $\Ads$ brane. 
Our approach to deal with this singularity is to peal off a small amount $n\ll p$ of $\Ads$ branes and consider them to be a probes in the backround of $p$ $\Ads$ backreacting branes. In that way the singularities in the warpfactor and the gauge fields, generated by the backreacting stack (see picture \ref{stack}), cancel each other since the singular contribution in the WZ term is of the same kind as the singular contribution in the DBI term. So this approach uses that string theory branes, even when non-supersymmetric, are locally behaving like supersymmetric branes where tension equals charge.  
This second approach effectively eliminates the diverging contributions of the geometry and the gauge field that couples to the brane, but in our case, it does not take care of the singularity in $H_3$ (and hence $\lambda$). Since the singularity in $\lambda$ occurs at a place where the space-time metric is not to be trusted due to high curvature one has to be careful to conclude that the infinite flux-clumping destabilizes the system due to direct brane-flux decay. We therefore cut off the solution a string-length away from the singularities and investigate how large the flux-clumping is. In this section we find that the flux-clumping $\lambda$ is proportional to $g_sp$. We also show that for large $g_sp$ the system is destabilised independent of the ratio $p/M$ being very small. 

Another systematic way to approach the effective field theory would be to consider high-order corrections in the blackfold approach \cite{Emparan:2011hg, Armas:2016mes}, which we hope to adress in the future. 

\subsection{Flux-clumping at the cut-off}
To evaluate the flux-clumping at a finite distance $r=\ell_s$ from the antibranes, we need to evaluate $\lambda$ there. Here we work directly with the $\Ads$ solution but later when we consider the T-dual background where the expression for $\lambda$ remains the same.  
Near a stack of extremal antibranes we have
\begin{equation}
\e^{-4A} = \e^{4B} = h_6~,\qquad \e^{\phi} = g_s h_6^{-\frac{3}{4}}\,,
\end{equation} 
with
\begin{equation}
 h_6(r) = \frac{g_s p ~\ell_s }{2 r} \,. \label{D6warpfunction}
\end{equation}
Since
$\lambda = \alpha \e^{\phi-7A}$, we can calculate $\lambda$ once we know the value of $\alpha$ near the cut-of. The latter we obtain from the Smarr-like relations found in \cite{Gautason:2013zw, Cohen-Maldonado:2015ssa, Cohen-Maldonado:2016cjh} (see also \cite{Ferrari:2016vcl}), that relate the energy $E$ (above the supersymmetric vacuum), to the values of the gauge fields at the sources:
\be \label{Smarr}
E = p\mu_6\int_7 C_7\,,\qquad C_7 = \alpha \d x_0\wedge\ldots\wedge\d x_6\,,
\ee
such that $E=\mu_6 p\alpha(r=0)$. To find $\alpha(0)$ we therefore need to know $E$. Following the reasoning of \cite{Maldacena:2001pb} we have (the background has $e^A=1$ at the tip)
\be
E=2\mu_6 p g_s^{-1}\,,
\ee
From the Smarr relation (\ref{Smarr}) we then find
\be
\alpha(0) = 2g_s^{3/4}\,. 
\ee
Concerning $\lambda$ we have $\lambda(l_s) = g_s^{1/4}\frac{\pi}{2}p \alpha(l_s)$. We take $\alpha(l_s)=\alpha(0)$  and find  $\lambda \sim g_s p$, consistent with the estimates in \cite{Michel:2014lva}. One can only consistently take $\alpha(l_s)=\alpha(0)$ if $\alpha'(0)$ is sufficiently small but this is easy to see.
At the brane source we must have $F_2 = Q~\Omega_2$ and this provides an expression for $\alpha'(0)$. Near the source we have: 
\be
F_2=g_s^{-\tfrac{3}{2}}(g_sp)^2\alpha'(0)\Omega_2\,.
\ee
such that
\be
\alpha'(0)=\frac{g_s^{1/2}}{g_sp}\ll 1\,.
\ee

\subsection{The corrected probe result at large $p$}\label{sec:corrected}
We now compute the probe potential for $n$ $\Ads$-branes near the tip geometry with the backreacting of $p$ $\Ads$-branes included. By cutting of the divergent flux-clumping by the value found above: $\lambda \sim g_s p$ we find a finite and well-defined probe potential. To carry out this computation we need the solution T-dual to the $\Ads$ solution. A straightforward computation gives the following metric for the smeared $\Adv$-branes at $r=0$:
\bea
\dd s^2&=& \e^{2A}\dd s_6^2 + \e^{-2A}\dd\psi^2 + \e^{2B} \left(\dd r^2 + r^2 \dd \Omega_2^2\right)~,\\
F_1 &=& (M/\ell_s)\dd\psi~,\\
F_3 &=& \e^{6A}\star_4 \dd \alpha~,\\
H_3 &=& \lambda\e^{\phi+A}\star_4 F_1~,\\
\e^{\phi_5} &=& \e^{\phi-A}~.
\eea
Where as before, the equations of motion demand
$\alpha=\lambda\e^{7A-\phi}$
and the fields have the standard D-brane singularities describing smeared 5-branes at $r\to 0$. We now follow the exact same procedure of Appendix  \ref{App1:potential} to compute the probe potential by placing a single NS5/$\overline{\text{NS5}}$-pair in the above background. This calculation can be found in Appendix \ref{App3:potential}.  The result is a run-away potential:
\be
V\propto g_s p\left(\f{n}{M}-\f{\Delta\psi}{\ell_s}\right)~.
\ee
We note that  the interaction term has a different dependence on the dilaton and warpfactor such that the warpfactor and dilaton do not simply factor to the front. The factor  $\e^{3A -\phi}$ in front of the interaction term reduces to $g_s^{-1}$ close to the brane, consistent with the uncorrected probe potential. 

Note that this potential receives large corrections as $\psi$ grows. The reason is that the decay of the branes causes at the same time a materialization of $M-n$ $D6$-branes which quickly dominate over the $p$ $\Ads$-branes such that the original background  geometry cannot be trusted.  Essentially, corrections are subleading until the value of $\psi$ is such that the drop in energy is twice the tension of the $n$ $\Ads$-branes. It is satisfactory that at the value for $\psi$ where the induced 6-brane charge in the KK5/$\overline{\text{KK5}}$ pair becomes zero, the drop in energy is two times the tension of the $n$ $\Ads$-branes. 

For a stable configuration we would expect that by dialling down $n/M$ and $p/M$ a metastable state would appear for the $n$ mobile antibranes. This is however not the case which indicates that the whole stack of branes may be unstable. Of course when $p$ keeps decreasing and eventually reaches small values for which our analysis is not valid. This is the limit of our approach and a different analysis must be carried out to determine the stability of small number of antibranes.

\section{$\overline{\text{D}k}$-branes with $k<6$} \label{sec:p<6}
In this section  we discuss how simple dimensional analysis can be employed to study flux clumping around antibranes of different dimensions. 

The flux-clumping for unpolarised $\overline{\text{D}k}$-branes can be inferred either via the supergravity techniques of \cite{Gautason:2013zw, Blaback:2014tfa} or the brane effective field theory techniques of \cite{Michel:2014lva}. Both give the same result, namely
\be
\lambda \sim g_sp \frac{\ell_s^{7-k}}{r^{7-k}}\,,
\ee
with $r$ a local radial variable. Hence the flux clumping at the cut-off scale $r=\ell_s$ always equals
\be
\lambda_c\sim g_s p\,.
\ee
In the supergravity regime this is too large to ensure stability of the probe. Now the question arises how $\lambda_c$ changes when the branes polarise via the Myers effect as predicted by KPV \cite{Kachru:2002gs}. The essential observation of this paper is that $\lambda_c$ does not change for $\Ads$-branes polarising into a KK5/$\overline{\text{KK5}}$ pair, because the pair has the same dimensionality as the original branes. We expect similar conclusion for $\Adv$-branes and hence they should be unstable as well in the supergravity regime. However when $k<5$ the $\overline{\text{D}k}$ charge spreads out on a spherical NS5-brane of radius $R$ such that one expects, from dimensional analysis,
\be
\lambda \sim g_sp\frac{\ell_s^{7-k}}{R^{5-k}r^{2}}\qquad \Rightarrow\qquad  \lambda_c \sim g_sp\frac{\ell_s^{5-k}}{R^{5-k}}\,. 
\ee
The largest possible value for $R$ is the size of the full A-cycle, which we take proportional to $R_{\text{max}}^2\sim g_sM\ell_s^2$ like in the KS throat. For $\overline{\text{D}4}$-branes we then have that this minimal value of $\lambda_c$ is still too large. However for $\Adt$-branes the situation improves. The radius at which the clumping is only of order 1 is found to be:
\be
R\sim \sqrt{g_sp}~\ell_s\,.
\ee 
This is far off the radius one finds from the probe computation, which is linear in $p$ \cite{Kachru:2002gs}:
\be
R_{\text{probe}} \sim \frac{p}{M}\sqrt{g_sM}~\ell_s\,.
\ee 
In fact at the probe radius the clumping can be estimated to be too high and should destabilise the brane. This already shows that the probe computation at best only gives qualitative information but certainly not quantitative results, very much in contrast with supersymmetric vacua \cite{Bena:2008dw}. Interestingly in reference \cite{Cohen-Maldonado:2015ssa} a $\sqrt{p}$ dependence of the polarisation radius was inferred on very different grounds: namely by using the aforementioned boundary condition for polarised branes that can regularise $\lambda$ without the need to cut-off. This potential match seems interesting and we leave it for future research.

\section{Conclusions} \label{concl}
We now summarise our results and then give an outlook on the metastability of general antibranes.

\subsection{Summary}

We have shown that at probe level $\Ads$-branes experience a classical barrier against annihilation with fluxes of the opposite charge on the condition that the Romans mass quantum $M$ is much bigger than the antibrane charge $p$, thereby extending the results of \cite{Gautason:2015tla}. This condition is (qualitatively) T-dual to the condition for the metastability of $\Adt$-branes \cite{Kachru:2002gs}, where Romans mass gets replaced with RR 3-form flux.  We have then shown that backreaction washes away the classical barrier, confirming the arguments of \cite{Blaback:2012nf} and \cite{Danielsson:2014yga}.

The reason this could be firmly shown rests on 2 facts:
\begin{enumerate}
\item $\Ads$-branes are special in the sense that they brane-flux decay via mutating into a KK5/$\overline{\text{KK}5}$ pair carrying $\Ads$-charge. The pair has effectively the same dimensionality and backreaction as localised $\Ads$-branes.
\item The backreaction of localised $\Ads$-branes is described by ODE's. Near the source the backreacted geometries are T-dual to $\overline{\text{D}k}$-branes with $k<6$ smeared over the $k-6$ directions inside the $A$-cycle that is filled with RR-flux. For such branes there was already strong evidence that they are unstable \cite{Gautason:2015ola}.
\end{enumerate}

The vanishing of the classical barrier upon backreaction is due to the attraction of the fluxes towards the $\Ads$-branes, something that is not taken into account in the probe approximation. This increase in flux diverges close to the antibranes at the classical level. Hence we followed the procedure of \cite{Goldberger:2001tn, Michel:2014lva}, which amounts to cutting off the singularity at string scale. The main observation is that the cut-off value for the flux density is still too high and causes immediate brane-flux decay. Effectively the increased flux density causes an increase in the probability that actual D6-branes materialise out of the flux cloud which subsequently annihilate with the $\Ads$-branes. Our result explains the lack of smooth $\Ads$ solutions with finite temperature \cite{Bena:2013hr}. 

The reason for this apparent failure of the probe approximation could be due to the fact that the probe calculation is outside of the regime of validity of the NS5 (or KK5) brane action.

We have also investigated whether the same picture could still be valid for $\overline{\text{D}k}$-branes with $k<6$. Using dimensional analysis, similar to \cite{Michel:2014lva} we argue that the same physics should hold for $\overline{\text{D}4}$ and $\overline{\text{D}5}$ branes but that $\Adt$-branes cannot be shown to be unstable using the method described in this paper. We do however find that, if a metastable state exists for $\Adt$-branes, the radius of the spherical NS5-brane must scale different with the charge $p$ from what is  predicted by the probe calculation \cite{Kachru:2002gs}. 

\subsection{Outlook}
It has been argued that the instability we find can be \emph{circumvented} by considering small brane charges (e.~g.~$p=1$) \cite{Michel:2014lva,Polchinski:2015bea}. For small brane charges the cut-off value of the flux clumping is small and furthermore there is no polarisation potential that can be computed within the supergravity regime. Given that flux clumping is not severe for small $p$, an instability cannot be argued solely based on a singular flux density. Instead brane effective field theory should be employed to argue for stability \cite{Michel:2014lva}. Stability would most likely imply a finite temperature solution with small $p$ in supergravity. However for $\Ads$-branes the absence of a smooth finite temperature solution was demonstrated in \cite{Bena:2013hr}. It could of course be that the metastable state at $p=1$ has a very small gap such that the minimal temperature needed for a classical horizon already destabilises the vacuum \cite{Polchinski:2015bea}. This potential loophole might become visible by computing finite temperature corrections to probe actions, something that can be done in the blackfold approach \cite{Emparan:2011hg, Armas:2016mes}.

Despite the possible room for the existence of metastable $\Adt$-branes, we want to be open-minded and contemplate about the possibility that also $\Adt$-branes are unstable against the flux-clumping effect \cite{Blaback:2012nf, Danielsson:2014yga}. As long as numerical evidence showing that the boundary conditions for regular solutions described in \cite{Cohen-Maldonado:2015ssa} and \cite{Cohen-Maldonado:2016cjh} can be taken, does not exist, the option of an instability has to be taken seriously. There exists a heuristic interpretation of the ``Smarr relation'' for flux throats that supports this interpretation. 

The generalised ADM mass $M_{ADM}$ of a flux throat with sources, measures the energy above the supersymmetric vacuum. It was shown that $M_{ADM}$ obeys a Smarr-like relation \cite{Cohen-Maldonado:2015ssa, Cohen-Maldonado:2015lyb, Cohen-Maldonado:2016cjh}\footnote{We have mentioned several times that a boundary condition for the gauge fields exists that allows a solution without infinite flux clumping. This boundary condition is simply $\Phi_M=0$ and cannot exist for $\Ads$-branes as shown in \cite{Blaback:2011nz}.}
\be
M_{ADM} =  Q_M\Phi_M + Q_D\Phi_D\,,
\ee 
where $Q_M$ ($Q_D$) stands for monopole (dipole) charge and $\Phi_M$ ($\Phi_D$) denotes the gauge field sources by the monopole (dipole) charge evaluated at the horizon in a specific gauge. The exact expression, on the nose, corresponds to the on-shell Wess-Zumino action for the would-be polarised brane (see \cite{Gautason:2013zw} for a similar derivation using on-shell brane actions). This almost fits the standard expectation that the energy of the flux throat equals twice the antibrane tension since
\be
M_{\text{ADM}} =  \text{DBI} + \text{WZ}\,,
\ee 
where  the DBI-term and WZ-term exactly equal each other in the probe approximation. In the backreacted case it seems all energy is therefore carried by the WZ term alone. This comes about since the warpfactor redshifts the on-shell DBI to zero, and alternatively this can be seen as a flux clumping effect, which makes the WZ dominate over the DBI. If this interpretation is correct then a metastable state can never exist since metastability truly requires a competition between the DBI and WZ term. We do want to point out that this is not a proof since our interpretation of the Smarr relation as on-shell probe potential is heuristic.

Finally, we have not touched upon the 3-form singularities in those (supersymmetric) AdS$_7$ solutions of \cite{Apruzzi:2013yva} with pure $\Ads$-branes (ie not all 6-brane charges come from spherical D8-branes). The logic of this paper suggests that those AdS$_7$ backgrounds, despite supersymmetry, are unstable and not suitable candidates for holographic descriptions of certain 6D CFT's. The typical 6D CFT's classified by these massive IIA solutions are however described mainly in terms of spherical 8-branes \cite{Gaiotto:2014lca}, but it would be interesting to understand this better.

\section*{Acknowledgements}
We thank Brecht Truijen for contributions to this project at an early stage.  We also acknowledge useful emails with Joe Polchinski on the backreaction of polarised branes and stimulating discussions with Iosif Bena, Gavin Hartnett, Bert Vercnocke and Herman Verlinde. We thank Marjorie Schillo for corrections in version 2. UD is supported by the Swedish Research Council (VR).   The work of FFG was supported by the John Templeton Foundation Grant 48222. The work of TVR is supported by the FWO odysseus grant G.0.E52.14N. We furthermore acknowledge support from the European Science Foundation Holograv Network.

\appendix

\section{O6 planes in flat space}\label{flatO6}
The O6-plane metric is given by
\be
ds^2= \f{1}{\sqrt{h_\text{O6}}} \dd s_7^2 + \sqrt{h_\text{O6}}\left(\dd r^2 + r^2 \dd\Omega_2^2\right)~,
\ee
where
\be
h_\text{O6} = 1-\f{g_s\ell_s}{4\pi r}~.
\ee
We expand this metric around the critical radius
\be
r = r_c + \delta r = \f{g_s \ell_s}{4\pi}+ \delta r~.
\ee
The result is
\be
\dd s^2 \approx \sqrt{\f{g_s \ell_s}{4\pi \delta r}} \dd s_7^2 + \sqrt{\f{4\pi\delta r}{g_s \ell_s}}\left(\dd \delta r^2 + \left(\f{g_s \ell_s}{4\pi}\right)^2 \dd \Omega_2^2\right)~.
\ee
By changing coordinates 
\be
x \equiv \f45 \left(\f{4\pi}{g_s \ell_s}\right)^{1/4} \delta r^{5/4}~,
\ee
we obtain the final expression 
\be
\dd s^2 \approx \left(\f{g_s \ell_s}{5\pi x}\right)^{2/5}\dd s_7^2 +\dd x^2 + \left(\f{g_s \ell_s}{4 \pi}\right)^2\left(\f{5\pi x}{g_s \ell_s}\right)^{2/5}\dd \Omega_2^2~.
\ee
 
\section{Computing the probe potential}\label{App1:potential}
We now turn to calculating the probe potential.  In order to obtain the expression for $B_6$ we make use of the Bianchi identity $\dd\big(H_7 - F_1 C_6\big)=0$, from which we infer
\be
H_7 = \dd B_6 +  F_1 C_6~.
\ee
The RR gauge potential $C_6$ can be calculated from the expression for $F_7=\dd C_6$ as
\be
C_6 = g_s^{-1}(S^{-1}+c)\vol_6~,
\ee
where the constant $c$ explicitly shows the gauge freedom in the definition of $C_6$. Using this we find
\be
B_6 =  g_s^{-1}\f{cM}{\ell_s}\psi\vol_6~.
\ee
The WZ terms in \eqref{braneaction} can be evaluated to
\be
\mu_5 g_s^{-1}M S^{-1}\left(\f{\psi_0}{\ell_s} - \f{p}{2M}  - \f{p}{2M}cS\right)~.
\ee
We could worry about the last term in the above expression which seems to break the gauge invariance of the potential. It is due to the fact that $\delta B_6 = - C_0\delta C_6$ and
\be
\delta (B_6-{\cal G}_0 C_6) = \f{p}{2}\delta C_6~.
\ee
The WZ Lagrangian is therefore not gauge invariant as we have written it down. One can introduce an extra worldvolume field to cure this problem. We will not pursue this direction since the term in question represents an overall shift in the potential and can be used to set our zero-point of the potential.

Turning to the DBI action, we find
\be
-\mu_5 g_s^{-2} S^{-1/2}\sqrt{1+g_s^2M^2 S^{-1}\left(\f{p}{2M} - \f{\psi_0}{\ell_s} \right)^2}~.
\ee
Using the expression \eqref{Seq} for $S$ and evaluating at the tip $r=0$ we obtain the full probe action 
\be
S_\text{NS5} =  -\mu_5 g_s^{-1}M v^{-2}\left(\left|\f{p}{2M} - \f{\psi_0}{\ell_s} \right|-\f{\psi_0}{\ell_s} +\f{p}{2M}  + \f{p}{2M}c v^2\right),
\ee
where we used that $v \ll g_s M$ and disregarded terms of order $(v/M g_s)^2$. Later we will add an interaction term which is of order $v/M g_s$ and is therefore leading order with respect to the terms we ignored.
So far we have calculated the probe action of the NS5-brane located at $\psi_0$. We also have to add the action of the $\overline{\text{NS5}}$-brane located at $-\psi_0$. This brane has opposite charge and sign of $p$ as explained above. The combined potential is then the same as above, but with an overall factor of 2:
\be
V(\Delta\psi) = \mu_5 g_s^{-1}M v^{-2}\left(\left|\f{p}{M} - \f{\Delta\psi}{\ell_s} \right|-\f{\Delta\psi}{\ell_s} +\f{p}{M}\right)~,
\ee
where $\Delta\psi = 2\psi_0$ is the separation of the two branes.
Here we have also fixed the zero-point energy by taking $c=0$ such that at $\psi=1/2$ the potential vanishes.

Notice that for $\Delta\psi \ge \ell_s p/M$ the potential vanishes, this is because we have not taken into account the electromagnetic and gravitational interaction between the two branes. For a brane-antibrane pair this is a subleading effect but is obviously important to obtain the full potential. We can compute the interaction potential between the two probe NS5-branes by placing one NS5-brane in the geometry and let it backreact. This computation is spelled out in the next Appendix \ref{App2:potential}.

\section{The force between an NS5/$\overline{\text{KK}5}$ pair}\label{App2:potential}
We can compute the interaction potential between the two probe NS5-branes by placing one NS5-brane in the geometry, let it backreact, and evaluate that backreaction in the probe brane action of the other NS5-brane. Hence,
to find the force between an NS5/$\overline{\text{NS5}}$ pair amounts to solving a differential equation equation for the harmonic function, denoted $K$, describing a NS5-brane in our throat geometry. 

We are only interested in the probe potential and thereby the interaction term close to $r=0$, which simplifies our task immensely. The background metric at $r=0$ simplifies to a cylinder:
\be
ds^2 \to v^{-1} ds_6^2 + v(\dd \psi^2 + \dd r^2 + r^2 \dd \Omega_2^2)~.
\ee
A single NS5-brane in this background deforms the metric and fields in the standard way:
\bea
\dd s^2&=& v^{-1}\dd s_6^2+ vK \left(\dd \psi^2 + \dd r^2 + r^2 \dd \Omega_2^2\right)~,\\
H_3 &=&   \tilde\star_4 \dd K~,\\
H_7 &=&  g_s^{-2}v^{-2}\vol_6\w \dd K^{-1}~,\\
\e^{\phi}&=& g_s v^{-1}K^{1/2}~,
\eea
The form equations of motion reduce to
\be
\dd \tilde\star_4 \dd K = 0~,
\ee
with the solution 
\be
K = 1 + \f{4\pi^3 \ell_s}{r}\f{\sinh (2\pi r/\ell_s)}{\cosh (2\pi r/\ell_s) - \cos (2\pi\Delta\psi/\ell_s)}~,
\ee
where we have put the NS5-brane at the location $r=0$ and $\Delta\psi$ measures the distance from the brane along the isometry direction. Furthermore we have put the number of NS5-branes to one and used that in our conventions $\mu_5 = 2\pi$.
In the region around $r= 0$ we have 
\be
B_6 = \f{1}{v^2 g_s^2} \f{\sin^2 \left(\pi \Delta \psi/\ell_s\right)}{4\pi^4 + \sin^2 \left(\pi \Delta \psi/\ell_s\right)}~.
\ee
We obtain the interaction term by inserting this into the action of the other NS5-brane so that we get
\be
-\mu_5 \int \left(\e^{-2\phi}\sqrt{-g} +B_6\right)= -2\f{\mu_5}{v^2 g_s^2}  \f{\sin^2 \left(\pi \Delta \psi/\ell_s\right)}{4\pi^4 + \sin^2 \left(\pi \Delta \psi/\ell_s\right)}\int\vol_6~.
\ee
The potential describing the force starts rising until it reaches a maximum and then goes back in an identical matter. In reference \cite{Gautason:2015tla} this potential was approximated to be piecewise linear, which is qualitatively similar. The reason the potential grows roughly linear with $\psi$ is due to the throat geometry; the brane/antibrane forces are effectively confined to the circle dimension such that it mimics electrodynamics in $1+1$ dimensions.

\section{Computing the corrected probe potential}\label{App3:potential}
To evaluate the probe potential of a single NS5/$\overline{\text{NS5}}$-pair in the background described in section \ref{sec:corrected} we require the gauge field expressions:
\bea
C_0 &=& (M/\ell_s)\psi~,\\
C_6 &=& (\alpha + g_s^{-1} c)\vol_6~,\\
B_6 &=& \f{c M}{\ell_s g_s} \psi \vol_6~.
\eea
Keeping in mind that NS5-brane pair carries $n$ units of antibrane charge, the WZ terms are
\be
\mu_5 g_s^{-1}M\left(-\f{\Delta\psi}{\ell_s} \alpha g_s + \f{n}{M} \alpha g_s - \f{n}{M} c\right)~.
\ee
In the same spirit the DBI action takes the form
\be
S_\text{DBI}=-\mu_5 M \e^{6A-\phi_5} \left| \f{n}{M} - \f{\Delta\psi}{\ell_s}\right|=-\mu_5 M \e^{7A-\phi} \left| \f{n}{M} - \f{\Delta\psi}{\ell_s}\right|~,
\ee
and combined
\be
V = \mu_5 M \e^{7A-\phi}\left(\left| \f{n}{M} - \f{\Delta\psi}{\ell_s}\right| - \f{\Delta\psi}{\ell_s} \lambda + \f{n}{M} \lambda + \f{n}{M} cg_s^{-1}\e^{-7A+\phi}\right)~. 
\ee
To calculate the interaction term we again follow the same procedure as in Appendix \ref{App2:potential}. The metric is approximated by the cylinder metric evaluated at a given radius $r=\ell_s$
\be
\dd s^2 = \e^{2A}\dd s_6^2  + \e^{2B} \left(\e^{-2B-2A}\dd\psi^2 + \dd r^2 + r^2 \dd \Omega_2^2\right)~,
\ee
We rescale $r$ to absorb the factors of $\e^{B-A}$ as $\tilde r = \e^{B-A}r$. Using the method described in Appendix \ref{App2:potential} we find the interaction potential 
\be
V_\text{interaction} = \mu_5M \e^{7A-\phi}\left[\f{2\e^{3A-\phi}}{M}\left(1 + \f{4\pi^3 \ell_s}{\tilde r}\f{\sinh (2\pi \tilde r/\ell_s)}{\cosh (2\pi \tilde r/\ell_s) - \cos (2\pi\Delta\psi/\ell_s)}\right)^{-1}\right]~.
\ee
In total we find the following correction of the probe result \ref{probeV1}:
\be
V = \mu_5 M \e^{7A-\phi}\left(\left| \f{n}{M} - \f{\Delta\psi}{\ell_s}\right| - \f{\Delta\psi}{\ell_s} \lambda + \f{n}{M} \lambda \right) + V_\text{interaction}~. 
\ee
where we threw away an additive constant. If we were to take  $\tilde{r}$ to zero we would find that $\lambda$ blows up in exactly the right way, see (\ref{lambdablowup}), to compensate for the vanishing prefactor, and hence we obtain finite result for the terms that include $\lambda$. All other terms, including the interaction term, vanish in this limit, which physically can be understood as an effect due to redshift. Since we cannot trust these expressions below the string scale, we need to evaluate them at $\tilde{r}=\ell_s$ to investagate whether the redshift is large enough within the regime under our control. In paricular, we find that $\lambda$ at its cut-off value is $\lambda=g_sp$, which, indeed, is large.

\bibliographystyle{utphys}
{\small
\begingroup
    \setlength{\bibsep}{6pt}
    \setstretch{1}
    \bibliography{refs}
\endgroup}
\end{document}